\documentclass[hyper]{JHEP3}
\usepackage{amssymb}
\usepackage{amsfonts}
\usepackage{amsbsy}
\usepackage{amsmath}
\usepackage{amssymb}
\usepackage{amsfonts}
\usepackage{amsbsy}
\usepackage{amsmath}

\title{New F-theory lifts}

\author{Andr\'es~Collinucci\\

Institute for Theoretical Physics, Vienna University of Technology, \\
Wiedner Hauptstr. 8-10, 1040 Vienna, Austria\\
}

\abstract{In this note, a procedure is developed to explicitly construct non-trivial \mbox{F-theory} lifts of perturbative IIB orientifold models on Calabi-Yau complete intersections in toric varieties. This procedure works on Calabi-Yau orientifolds where the involution coordinate can have arbitrary projective weight, as opposed to the well-known hypersurface cases where it has half the weight of the equation defining the CY threefold. This opens up the possibility of lifting more general setups, such as models that have O3-planes.
}

\begin{document}
\section{Introduction}
F-theory \cite{Vafa:1996xn} is a tool that encodes information about IIB seven-brane models geometrically in an intrinsically non-perturbative way. However, Sen's procedure \cite{Sen:1997gv,Sen:1996vd, Sen:1997kw} puts F-theory models directly into contact with perturbative IIB orientifold setups with D7-branes, by taking the weak coupling limit in a geometric way. This limit of Sen has allowed for much progress in understanding IIB vacua, as it makes it possible to relate F-theory models to M-theory and Heterotic compactifications. The F-theory/IIB link allows one to reliably compute many useful quantities such as curvature induced charges on D7's and O7's via the formula in \cite{Sethi:1996es}, the dimensions of moduli spaces of IIB setups, and to establish necessary conditions for D-instantons to make contributions to superpotentials \cite{Witten:1996bn}. For a thorough introduction to F-theory, see \cite{Denef:2008wq}

However, the reverse procedure of lifting a given IIB orientifold compactification on a CY threefold to an F-theory CY fourfold is not yet under full control. Although the procedure to do this is in principle clear, the technicality of the task has limited the literature to studying only a subset of models where the IIB Calabi-Yau threefold is a hypersurface given by an equation of the form
\begin{equation}
\xi^2+P=0\,, \label{senform}
\end{equation}
and the orientifold involution is $\xi \rightarrow -\xi$. Such models include the $\mathbb{P}_{1,1,1,6,9}(18)$ geometry studied in \cite{Denef:2004dm}, the $\mathbb{P}_{1,1,1,1,4}(8)$ and $\mathbb{P}_{1,1,2,2,6}(12)$ models used in \cite{Giryavets:2003vd}, and also the $\mathbb{P}_{1,1,1,2,5}(10)$ model. The corresponding fourfolds for some of those models can be found in \cite{Klemm:1996ts}. In all these cases, the last coordinate, whose degree is half the degree of the CY hypersurface, is taken as the involution coordinate, such that the hypersurface has the familiar form \eqref{senform}. Other models that have been succesfully studied are toroidal CY's, \cite{Denef:2005mm, Dasgupta:1997cd}.
This is a severe limitation on model building, as one cannot choose the involution at will, and cannot generate O3-planes. In fact, one cannot lift any O7/O3 models on the CY quintic! 

One possible obstruction to finding more general lifts is perhaps the fact that, until recently, the computation of the D3-tadpole induced by a generic involution invariant D7-brane on a divisor of a CY threefold was not possible. In principle, the formula by Sethi, Vafa and Witten in \cite{Sethi:1996es} relates the Euler characteristic of the F-theory fourfold to the induced D3 charges on the D7-brane and O7-plane. Numerically, however, this does not match up unless one takes into account the fact that a generic involution invariant D7-brane is singular. Therefore, even if one knew how to construct an F-theory fourfold for a given IIB setup, one had no reliable way of checking, whether it was the right one. 

In \cite{Collinucci:2008pf}, this mismatch was solved by taking the singularity of the D7-brane into account, thereby bridging the gap between F-theory and IIB theory. The main message of that paper, however, is the fact that D7-branes have a more natural habitat in the tachyon condensation picture of Sen, \cite{Sen:1998sm}, whereby in the orientifold case only even ranked stacks of D9/anti-D9 pairs are allowed. This is the technique that I will use in this note, as opposed to blowing up singularities.
\vskip 2mm
In this note, a simple procedure is presented to explicitly construct F-theory fourfolds (with sections) from arbitrary  IIB orientifold models on complete intersection Calabi-Yau threefolds with arbitrary, holomorphic orientifold involutions of type $h^{1,1}_-=0$. The basic idea is to first construct the base for an elliptic fibration, which will be the $\mathbb{Z}_2$ quotient of a CY threefold, as the complete intersection of $n-3$ equations in an $n$-dimensional toric manifold. This is in contrast to the usual known cases, where the base is itself already toric. Therefore, the F-theory fourfold will end up being the complete intersection of those $n-3$ equations with the Weierstrass equation in a toric $n+1$-fold. The procedure will be worked out in detail for the Calabi-Yau quintic, where an O3-plane is present; and in a case without O3's, in the resolved $\mathbb{P}_{1,1,2,2,2}(8)$ hypersurface.

\section{The quintic permutation orientifold}
\subsection{IIB data}
\subsubsection{The CY orientifold}

In this section, we will consider a IIB orientifold based on the permutation of two coordinates of the quintic Calabi-Yau manifold. Defining the projective space $\mathbb{CP}^4$ with the homogeneous coordinates $(x_1: x_2: x_3: x_4: x_5)$, the quintic CY threefold is defined as the vanishing locus of a homogeneous degree five polynomial.

To setup a IIB orientifold theory of type O7/O3 on this space, we need to define a holomorphic involution, i.e. a holomorphic map from the space to itself that squares to the identity:
\begin{equation}
\sigma: X_3 \rightarrow X_3 \quad {\rm such\ that} \quad \sigma \circ \sigma={\rm id}\,.
\end{equation}
We will choose $\sigma$ to be the map $x_5 \rightarrow -x_5$. Via a linear coordinate transformation, this involution is equivalent to a permutation of two coordinates, however, it is easier to work with the choice made here. In order for the CY threefold to be invariant under this involution, it is sufficient to require that the defining quintic polynomial contain only even powers of $x_5$. As long as there is a term of the form $(x_5)^4\cdot P^{(1)}(x_1, \ldots, x_4)$, the polynomial will be transversal.

This involution has two fixed-point loci: 
\begin{enumerate}
\item {\bf O7-plane}: This is the divisor defined by $x_5=0$.
\item {\bf O3-plane}: This is the point defined by $x_1 = \ldots = x_4=0$, and $x_5=1$.
\end{enumerate}
Note, that this O3-plane \emph{is} contained in the quintic hypersurface, as we have banned the term ${x_5}^5$ from the quintic polynomial. 

\subsubsection{D$7$-brane geometry}
Since the involution has an O7-plane defined by $x_5=0$, which is a divisor of the hyperplane class $H$ (i.e. Poincar\'e dual to the generator of $H^2(X_3, \mathbb{Z})$), in order to cancel the D$7$ tadpole, there must be a D$7$-brane wrapped on a divisor $S$ of class $8\,H$. Following the Dirac quantization and the K-theoretic arguments put forward in \cite{Collinucci:2008pf}, we deduce that $S$ must be a singular surface mimicking the \emph{Whitney umbrella}, given by an equation of the following form:
\begin{equation}
\eta^2+x_5^2\,\chi = 0\,, \label{whitney1}
\end{equation}
where $\eta$ and $\chi$ are polynomials of degree $4$ and $6$, respectively.

As explained in \cite{Collinucci:2008pf}, there are two ways to deal with D7-branes of \emph{Whitney-type}: One can blow-up the double curve singularity given by the locus where the D7 intersects the O7, or one can contruct the D7 as the tachyon condensate of a rank two\footnote{It can be shown by a probe argument, \cite{Uranga:2000xp}, that one must in general have even-ranked stacks in order to cancel K-theory charges.} stack of D9-branes and its orientifold image anti-D9 stack. As is emphasized in that paper, the tachyon condensation picture is the most practical one. From it, we can compute the induced D$3$-charge almost instantly. 

Let us choose the gauge bundles on the stacks as follows:
\begin{center}
    \begin{tabular}{lcl}
      $\overline{{\rm D}9_1}\, \overline{{\rm D}9_2}\,,$ & \qquad & ${\rm D}9_1\, {\rm D}9_2$ \\
      $\overline{E}=\mathcal{O}(-a) \oplus \mathcal{O}(a-4)$\,,   & \qquad & $E=\mathcal{O}(a) \oplus \mathcal{O}(4-a)$\,, \\
    \end{tabular}
  \end{center}
where $a$ is some integer.
The tachyon field, $T$, is then a linear holomorphic bundle map $T:\overline{E} \mapsto E$. Alternatively, it is a section of $E \otimes E$. The orientifold condition on it is the following
\begin{equation}
\sigma^*(T) = -T^{\rm T}\,.
\end{equation}

The most general solution is represented by a two by two matrix of the following form
\begin{equation} 
 T = \left( \begin{array}{cc} 0 & \eta(u) \\ -\eta(u) & 0 \end{array} \right)
 + x_5 \, \left( \begin{array}{cc} \rho(u) & \psi(u) \\ \psi(u) & \tau(u) \end{array} \right)\,,
\end{equation}
where $(\eta, x_5, \rho, \psi, \tau)$ are of degree $\big(4, 1, 2\,a-1, 3, 7-2\,a\big) \,H$, respectively. In order for all entries to be globally well-defined sections, they must have positive degree, which means that we have the following bound:
\begin{equation}
\tfrac{7}{2} > a > \tfrac{1}{2}\,. \label{boundsquintic}
\end{equation}
The equation for the surface $S$ of the D7-brane will then be given by the locus where the tachyon map fails to be invertible, i.e. by the determinant equation
\begin{equation}
S: {\rm det}(T) = \eta^2-x_5^2\,(\psi^2-\rho\,\tau)=0\,.
\end{equation}
Note, that $(\psi^2-\rho\,\tau)$ is a non-generic form of the expected polynomial $\chi$ in \eqref{whitney1}. This restriction of the D$7$ geometric moduli signals the presence of a non-trivial U(1) flux on the D-brane. 

Let us now calculate the total charge vector, or \emph{Mukai} vector for this system, which is defined as the sum of the charge vectors of the D9 and anti-D9 stacks. The individual charge vectors are given by the well-known formula (\cite{Minasian:1997mm, Scrucca:1999uz, Stefanski:1998yx, Morales:1998ux}) ${\rm ch}(F)\,\sqrt{{\rm Td}(X_3)}$. The sum gives us
\begin{eqnarray}
\Gamma_{D7} &=& \Big(e^{a\,H}+e^{(4-a)\,H}-e^{-a\,H}+e^{(a-4)\,H}\Big)\,(1+\tfrac{c_2(X_3)}{24})\\
&=&8\,H+\Big(\frac{265}{3}+20\,(a-\frac{1}{2})\,(a-\frac{7}{2}) \Big)\,\omega\,,
\end{eqnarray}
where $\omega=H^3/5$, and we have used $c_2(X_3) = 10\,H^2$.
The O7-plane induces a D3 charge of $\chi(O7)/6=55/6$. The total induced D$3$ charge $Q^i$ coming from the D$7$ and the O$7$ is then equal to
\begin{equation}
Q^i = \frac{195}{2}+20\,(a-\frac{1}{2})\,(a-\frac{7}{2}) \,. \label{inducedquintic}
\end{equation}
It can be shown that the second term corresponds to a contribution of the form $F^2/2$, from a necessary U(1) flux that compensates for a Freed-Witten anomaly. A heuristic argument for this is the observation that, if one artificially saturates the bounds in \eqref{boundsquintic} by setting $a$ to $1/2$ or $7/2$, then the polynomial $\rho$ or $\tau$ becomes a constant, respectively. In the case $a=1/2$ (without loss of generality), the determinantal equation of the surface $S$ becomes:
\begin{equation}
{\rm det}(T) = \eta^2-x_5^2\,(\psi^2-\rho) = \eta^2-x_5^2\,\chi\,,
\end{equation}
after a redefinition of $\rho$. This means that the D7-brane no longer has a flux constraining its geometric moduli. Since the second term in \eqref{inducedquintic} vanishes when the bound is artificially saturated, we see that it must be the contribution due to a flux, which means that $195/2$ is the curvature induced part of the charge. In fact, one can directly calculate the so-called \emph{orientifold Euler characteristic} of the singular D7-brane defined in \cite{Collinucci:2008pf} and see that it matches this number. This can be proven rigorously by first blowing-up the surface $S$, and then computing the flux explicitly, which is a more cumbersome approach.

  Hence, the curvature induced part of the charge $Q^c$ is equal to $195/2$, and the rest, $Q^{F}$, is flux induced. The O$3$-plane contributes with a charge $1/2$, so that the total D$3$ tadpole is
\begin{equation}
Q^i+Q_{O3} = Q^c+ Q_{O3}+ Q^{F}= 98+20\,(a-\frac{1}{2})\,(a-\frac{7}{2})\,. \label{resultquintic1}
\end{equation}

\subsection{F-theory lift}
The purpose of this note is to give an F-theory description of the IIB setup introduced in the previous subsection. The main task is to construct a Calabi-Yau fourfold $Y_4$ that encodes the IIB setup in its geometry. We will construct $Y_4$ by running Sen's procedure in reverse: First, we will parametrize the quotient space $B_3 \equiv X_3/\mathbb{Z}_2$ of the orientifold in a convenient way. Second, we will proceed to construct an elliptic fibration over this quotient space that encodes the D$7$-brane and O7-plane data.\\

Define the weighted projective space $W\mathbb{CP}^4_{1,1,1,1,2}$ with corresponding coordinates\\ $(y_1, y_2, y_3, y_4, h)$. This will be the ambient space in which $B_3$ will be defined as a hypersurface. Now define the following map:
\begin{eqnarray}
q&:& W\mathbb{CP}^4 \rightarrow W\mathbb{CP}^4_{1,1,1,1,2}\\
q&:& (x_1, x_2, x_3, x_4, x_5) \mapsto (y_1, y_2, y_3, y_4, h) = (x_1, x_2, x_3, x_4, x_5^2)\,.
\end{eqnarray}
Notice that this map is everywhere surjective, and $2$ to $1$ everywhere except at the fixed-point loci, $x_5=0$, and $(0, 0, 0, 0, 1)$. The latter will induce an orbifold singularity in the quotient space. Our desired quotient space $B_3$ is defined as the hypersurface in the weighted projective space given by original quintic equation $P^5$, where ${x_5}^2$ gets replaced by $h$, and $x_i$ by $y_i$, for $i=1,2,3,4$:
\begin{equation}
\tilde P^{(5)} (y_1, y_2, y_3, y_4, h) \equiv P^{(5)} (y_1, y_2, y_3, y_4, h)\,, \label{quinticeq}
\end{equation}
whereby `$5$' is meant as the homogeneous degree of both polynomials, not as the power of the arguments, since $h$ has degree two.
This is possible because we required $P^{(5)}$ to depend only on even powers of $x_5$. 

We may now proceed to construct the elliptic fibration over $B_3$. In order for the fourfold to admit a weak coupling limit, we want a fibration of type $E_8$. The elliptic fiber is then a hypersurface in $W\mathbb{CP}^2_{2,3,1}$, with coordinates $(x,y,z)$, given by an equation of the form
\begin{equation}
y^2 = x^3 + f\,x z^4 + g\, z^6\,. \label{ellipticcurve}
\end{equation}
To turn this into a fibration over $B_3$, we must promote the coefficients $f$ and $g$ to polynomials in the base coordinates, and to make contact with Sen's orientifold limit of F-theory \cite{Sen:1997gv, Sen:1996vd}, these polynomials must be parametrized (redundantly) in the following way:
\begin{eqnarray}
f &=& -3\,h^2+C\,\eta\,,\nonumber\\
g &=& -2\,h^3+C\, h \, \eta- \frac{C^2\,\chi}{12}\,, \label{senvars}
\end{eqnarray}
where $h$ is the coordinate of weight two, $\eta$ and $\chi$ are the polynomials of degree $4$ and $6$ defined in the previous subsection, and $C$ is a constant that is taken to be small in the weak coupling limit.

In order for equation \eqref{ellipticcurve} to be well-defined, $(x,y,z)$ should also transform non-trivially under projective equivalence of the base space. The toric weights for the full ambient space are given in table \ref{tab:chargesfirstlift}.

\begin{table}[h!]
\begin{centering}
\begin{tabular}{|c|c|c|c|c|c|c|c|}
\hline 
$y_1$ & $y_2$ & $y_3$ & $y_4$& $h$ & $x$ & $y$ & $z$ \tabularnewline
\hline
\hline
1 & 1  & 1  & 1  & 2  & 2 & 3 & 0\tabularnewline
\hline
0 & 0  & 0  & 0  & 0  & 2 & 3 & 1\tabularnewline
\hline
\end{tabular}
\par\end{centering}

\caption{Projective weights under the two $\mathbb{C}^*$ actions for the ambient sixfold $Y_6$.}

\label{tab:chargesfirstlift}
\end{table} 
\newpage
Notice that we have defined a toric ambient \emph{sixfold}, $Y_6$, and the CY fourfold $Y_4$ will be given as a codimension two space defined by the intersection of the vanishing loci of the equations \eqref{quinticeq} and \eqref{ellipticcurve}. The Stanley-Reisner ideal for this toric space has the following two elements:
\begin{equation}
\{y_1 y_2 y_3 y_4 h\,\,; \, x y z \}\,.
\end{equation}
In this notation, each entry consists of a set of coordinates that are not allowed to vanish simultaneously.

We identify two inequivalent divisor classes:
\begin{eqnarray}
H &\equiv& D_{y_1} = D_{y_2}= D_{y_3} \\
F &\equiv& D_{y_1}+D_z\,.
\end{eqnarray}
In this basis we have:
\begin{equation}
D_h=2\,H\,, \quad \tfrac{1}{2}\,D_x = \tfrac{1}{3}\,D_y = F\,, \quad D_z = F-H\,.
\end{equation}
The Calabi-Yau fourfold $Y_4$ is defined as the intersection of two polynomials of class $5\,H$ and $6\,F$, respectively. The \emph{non-vanishing} intersection numbers are the following:
\begin{equation}
H^3\,F=H^2\,F^2=H\,F^3\,=F^4=\frac{5}{2}\,. \label{intfirstlift}
\end{equation}
The half-intergral nature of these numbers signals a $\mathbb{Z}_2$-orbifold singularity in the fourfold, due to the O3-plane. In the fourfold, the singularity is located at
\begin{equation}
\{y_1=0\} \cap \ldots \cap \{y_4=0\} \cap \{y^2 = x^3 + f\,x z^4 + g\, z^6\}\,, \label{o3quinticorbifold}
\end{equation}
which is just an elliptic curve in the fourfold. More specifically, it is the fiber above the locus of the O3-plane. Hence, in this F-theory lift, the O3 plane is encoded as an elliptic curve of $\mathbb{Z}_2$ singularities. Such singularities were described for toroidal fourfolds in $\cite{Dasgupta:1997cd}$, where it was shown that they are `terminal', i.e. do not admit a crepant resolution. They were also encountered in \cite{Denef:2005mm}. Since the O3-plane is not charged under the axio-dilaton, we should not expect the fiber to degenerate above it. This can be easily confirmed as follows. By inspecting the polynomials $f$ and $g$ in Sen's limit \eqref{senvars}, we see that they must be sections of $4\,H$ and $6\,H$, respectively. Evaluating the functions at the locus of the O3 \eqref{o3quinticorbifold}, we see that only monomials depending exclusively on $h$ can survive. Hence, the polynomials will be of the form:
\begin{equation}
f = a\,h^2\,, \quad {\rm and} \quad g = b\,h^3\,,
\end{equation}
where, $a$ and $b$ are constants that get contributions from all terms in \eqref{senvars}, not just the first terms. We can now compute the discriminant of the elliptic curve above this locus as follows:
\begin{equation}
\Delta = 4\,f^3+27\,g^2 = (4\,a^3+27\,b^2)\,h^6\,.
\end{equation}
This is non-zero for generic choices of $f$ and $g$. \emph{Hence, the elliptic fiber is perfectly regular above the O3-plane}.\\

Let us now perform a test on this F-theory lift. According to the formula in \cite{Sethi:1996es}, the Euler characteristic of $Y_4$ divided by $12$ should give us the total D3 tadpole as measured in the covering space of the orientifold. Using the adjunction formula, we can compute the total Chern class of the fourfold as follows:
\begin{equation}
c(Y_4) = \frac{(1+2\,H)\,(1+H)^4\,(1+2\,F)\,(1+3\,F)\,(1+F-H)}{(1+6\,F)\,(1+5\,H)}\,.
\end{equation}
The fourth Chern class is then
\begin{equation}
c_4(Y_4) = 360 F^4+12 F^3\,H+87\,F^2\,H^2+9\,F\,H^3+183\,H^4\,.
\end{equation}
From this, and the intersection numbers in \eqref{intfirstlift} we can compute the Euler number of $Y_4$
\begin{equation}
\chi(Y_4)=1170\,.
\end{equation}
Note, that we have neglected the orbifold singularity in computing this number. It is very well possible that the top Chern class of the tangent bundle `misses' this singularity due to its high codimension (in this case three). 

Now, we can compute the predicted value of the curvature induced D$3$ charge, as measured from the threefold covering space, from F-theory as follows:
\begin{equation}
\fbox{$\chi(Y_4)/12 = 195/2 = Q^{c}$}\,,
\end{equation}
which exactly matches the expected value for the curvature induced charge in \eqref{inducedquintic}. This calculation does not `see' the contribution from the O3-plane.

\section{The resolved $W\mathbb{CP}^2_{1,1,2,2,2}(8)$ orientifold}
In this section, I would like to provide an example where no O3-planes, and hence no singularities, are present, to show that the procedure works and gives the right results when calculations are reliable.

\subsection{IIB data}
To define the toric fourfold $\overline{X_4}$ that will be the ambient space of our Calabi-Yau threefold, we begin by writing down the coordinates, and their respective charge assignments with respect to the two $\mathbb{C}^*$ toric actions in table \ref{tab:chargesoctic}.

\begin{table}[h!] 
\begin{centering}
\begin{tabular}{|c|c|c|c|c|c||c|}
\hline 
$x_{1}$ & $x_{2}$ & $x_{3}$ & $x_{4}$ & $x_{5}$ & $x_{6}$ & \emph{p}\tabularnewline
\hline
\hline 
1 & 1 & 2 & 2 & 2 & 0 & 8\tabularnewline
\hline 
0 & 0 & 1 & 1 & 1 & 1 & 4\tabularnewline
\hline
\end{tabular}
\par\end{centering}

\caption{Projective weights under the two $\mathbb{C}^*$ actions for the ambient fourfold $\overline{X_4}$. The last column denotes the degrees of a CY hypersurface.}

\label{tab:chargesoctic}
\end{table} 
\newpage
The Stanley-Reisner ideal reads
\[SR=\{x_1 x_2\,; \, x_3  x_4  x_5 x_6\}.\]
The Fermat-like equation defining the Calabi-Yau three-fold $X_3$ is the following:
\begin{equation}
(x_1^8+x_2^8)\,x_6^4+x_3^4+x_4^4+x_5^4=0\,.
\end{equation}
In the basis $H=D_{1}=D_2$; $G=D_{3}, D_4, D_5$, this defines a divisor in the class $[4\, G]$.
The non-vanishing triple intersection numbers are
\begin{equation}
H\,G^2=4\,, \quad G^3 = 8\,.
\end{equation}
Using the adjunction formula, we can deduce the second Chern class of $X_3$:
\begin{equation}
c_2(X_3) = \big(6\,G^2+2\,G\,H\big)\,.
\end{equation}

Let us now pick the involution $x_3 \rightarrow -x_3$. The fixed-point locus is the O7-plane defined by the divisor $x_3=0$, of class $G$. Notice that the fixed-point locus given by $x_4= \ldots = x_6=0$ does not intersect $X_3$.

Running through the same procedure as before, we build the D9 and anti-D9 stacks with fluxes as follows:
\begin{center}
    \begin{tabular}{lcl}
      $\overline{{\rm D}9_1}\, \overline{{\rm D}9_2}\,,$ & \qquad & ${\rm D}9_1\, {\rm D}9_2$ \\
      $\overline{E}=\mathcal{O}(-a\,G) \oplus \mathcal{O}((a-4)\,G)$\,,   & \qquad & $E=\mathcal{O}(a\,G) \oplus \mathcal{O}((4-a)\,G)$\,. \\
    \end{tabular}
  \end{center}
Once again, we have the bounds\footnote{In this case, it might be possible that setting $a=1/2$ actually corresponds to the splitting principle construction of a \emph{bona fide} rank two bundle on $X_3$. It would be interesting to use the techniques in \cite{Anderson:2008uw} to check this.} $7/2 \geq a \geq 1/2$. The induced D3 tadpole in this case is:
\begin{equation}
\frac{400}{3}+32\,(a-\tfrac{1}{2})\,(a-\tfrac{7}{2})\,.
\end{equation}
Adding the induced charge from the O7-plane, we get the following:
\begin{equation}
Q^i=\frac{400}{3}+\frac{64}{6}+32\,(a-\tfrac{1}{2})\,(a-\tfrac{7}{2})
=144+Q^F\,, \label{resultoctic1}
\end{equation}
where the last bit is again a flux induced charge.

\subsection{F-theory lift}
Following the procedure defined in the previous model, we first define a $\mathbb{Z}_2$ orbifold of the ambient fourfold by the following map:
\begin{eqnarray}
q&:& \overline{X_4} \mapsto \overline{X_4}/\mathbb{Z}_2\\
q&:& (x_3, x_1, x_2, x_4, x_5, x_6) \mapsto (h, y_1, y_2, y_4, y_5, y_6) = (x_3^2,  x_1, x_2, x_4, x_5, x_6)\,.
\end{eqnarray}
The toric space $\overline{X_4}/\mathbb{Z}_2$ is defined by the data in table \ref{tab:chargessecondorbifold} 

\begin{table}[h!]
\begin{centering}
\begin{tabular}{|c|c|c|c|c|c|c|c|c|c|}
\hline 
$h$ & $y_1$ & $y_2$ & $y_4$ & $y_5$ & $y_6$ \tabularnewline
\hline
\hline
4 & 1 & 1 & 2 & 2  & 0 \tabularnewline
\hline
2 & 0 & 0 & 1 & 1  & 1 \tabularnewline
\hline
\end{tabular}
\par\end{centering}

\caption{Projective weights under the three $\mathbb{C}^*$ actions for the ambient fourfold $\overline{X_4}/\mathbb{Z}_2$.}
\label{tab:chargessecondorbifold}
\end{table} 
and the SR ideal 
\begin{equation}
\{y_1 y_2\,; \, y_4 y_5 y_6 h \}\,.
\end{equation}
The base $B_3$ of our F-theory fourfold is then a hypersurface given by an equation of the form
\begin{equation} \label{hypersurfacesecondorbifold}
h^2+(y_1^8+y_2^8)\,y_6^4+y_4^4+y_5^4=0\,.
\end{equation}

The corresponding F-theory fourfold $Y_4$ is readily made by creating an ambient sixfold $Y_6$, of which the toric weights are defined in table \ref{tab:chargesecondlift}.
\begin{table}[h!]
\begin{centering}
\begin{tabular}{|c|c|c|c|c|c|c|c|c|c|}
\hline 
$h$ & $y_1$ & $y_2$ & $y_4$ & $y_5$ & $y_6$ & $x$ & $y$ & $z$\tabularnewline
\hline
\hline
4 & 1 & 1 & 2 & 2  & 0 & 4 & 6 & 0\tabularnewline
\hline
2 & 0 & 0 & 1 & 1  & 1 & 2 & 3 & 0\tabularnewline
\hline
0 & 0 & 0 & 0 & 0  & 0 & 2 & 3 & 1\tabularnewline
\hline
\end{tabular}
\par\end{centering}

\caption{Projective weights under the three $\mathbb{C}^*$ actions for the ambient sixfold $Y_6$.}

\label{tab:chargesecondlift}
\end{table} 

The Stanley-Reisner ideal reads:
\begin{equation}
\{y_1 y_2\,; \, y_4 y_5 y_6 h\,; \, x  y  z \}\,.
\end{equation}

We define the basis
\begin{equation}
H=D_{y_1}=D_{y_2}\,, \quad G = D_{y_4}\,, \quad F=\tfrac{1}{2}\,D_{x} = \tfrac{1}{3}\,D_{y}\,.
\end{equation}
The CY fourfold $Y_4$ is then the intersection of the Weierstrass equation with \eqref{hypersurfacesecondorbifold}. The non-vanishing intersection numbers are the following:
\begin{eqnarray}
F^4&=&F^3\,G=F^2\,G^2=F\,G^3=4\,\nonumber\\
F^3\,H&=&F^2\,G\,H=F\,G^2\,H=2\,.
\end{eqnarray}
The fourth Chern class of the fourfold $Y_4$ is the following:
\begin{eqnarray}
c_4(Y_4) &=& \nonumber\\
&+&\big(360\,F^4+12\,F^3\,G+43\,F^2\,G^2+5\,F\,G^3+22\,F^2\,G\,H+2\,F\,G^2\,H\big)\,. \nonumber
\end{eqnarray}
We can easily compute the Euler number to be the following:
\begin{equation}
\chi(Y_4) = c_4(Y_4)\cdot 6\,F \cdot 4\,G = 1728\,.
\end{equation}
We can now compute the D3 tadpole, as measured from the CY threefold covering space, to be
\begin{equation}
\frac{1728}{12} = 144\,,
\end{equation}
which precisely matches the result from \eqref{resultoctic1}. 
\begin{equation}
\fbox{$\chi(Y_4)/12 = 144 = Q^{c}$}\,,
\end{equation}

\section{Conclusions}
In this note, a simple procedure was devised to construct an elliptically fibered F-theory CY fourfold given a CY threefold with an O7/O3-type orientifold involution with $h^{1,1}_-=0$. The method was worked out in two examples: The quintic CY with an O3-plane, and the resolved $\mathbb{WCP}_{1,1,2,2,2}(6)$ hypersurface without O3-planes. In the first case, we saw that the O3-plane yields a regular elliptic curve worth of $\mathbb{Z}_2$-orbifold singularities in the fourfold. 
The second case yielded a smooth CY fourfold. In both cases, the curvature induced D3-tadpoles predicted by computing the Euler characteristics of the F-theory fourfolds were successfully matched with the IIB K-theoretic predictions. This is a highly non-trivial test of these fourfold constructions.

Although the case with the O3-plane yielded a singular F-theory fourfold, the na\"ive Euler characteristic of the fourfold seems to `miss' this curve of singularities, as it numerically matches the tadpole calculation performed in the IIB tachyon condensation picture. This might be due to the high codimension of the singularity. 

I have no proof that this phenomenon will always happen, however I have checked two other simple cases where this persists: the $\mathbb{WCP}_{1,1,1,2,5}(10)$ hypersurface, where this na\"ive computation of the fourfold Euler characteristic gives $\chi=378$, which matches the expected IIB tadpole if one again excludes the O3 contribution; and the resolved $\mathbb{WCP}_{1,1,1,6,9}(18)$ hypersurface, where, the fourfold has a na\"ive Euler characteristic $\chi=216$, again matching the K-theoretic result. In both cases, I took a coordinate of weight one as the involution coordinate.

The second case worked out in this note has no O3-planes, and hence no singularities in the fourfold. In this case, one can reliably match the IIB tadpole calculation to the Euler characteristic of the fourfold.

The method can be easily generalized to any CICY threefold. If the threefold is, for instance, a hypersurface, the idea is simply to construct the $\mathbb{Z}_2$-orbifold of the CY threefold as a hypersurface of a toric fourfold, as opposed to the usual construction where the orbifold itself is already toric. Then, a fibration over the base is easily made by adding three coordinates and imposing a Weierstrass equation. In the end, the CY fourfold is described as a codimension-two submanifold of a sixfold. This can of course be generalized to CY's that are CICY's with higher codimension.
\vskip 2mm
This technique lifts a road block in IIB model building, hopefully making it possible to answer interesting questions such as the role of O3-planes in F-theory, and the necessary conditions for instantons to contribute to superpotentials in model building. 

\section*{Acknowledgements}
I am grateful to R. Blumenhagen, F. Denef, M. Esole, A. Uranga, and T. Wyder for very useful discussions and exchanges.  This work was supported in part by the Austrian Research Funds FWF under grant number P19051-N16.

\bibliographystyle{JHEP}
\bibliography{reflift}

\providecommand{\href}[2]{#2}\begingroup\raggedright\begin{thebibliography}{10}

\bibitem{Vafa:1996xn}
C.~Vafa, {\it {Evidence for F-Theory}},  {\em Nucl. Phys.} {\bf B469} (1996)
  403--418, [\href{http://arxiv.org/abs/hep-th/9602022}{{\tt hep-th/9602022}}].

\bibitem{Sen:1997gv}
A.~Sen, {\it {Orientifold limit of F-theory vacua}},  {\em Phys. Rev.} {\bf
  D55} (1997) 7345--7349, [\href{http://arxiv.org/abs/hep-th/9702165}{{\tt
  hep-th/9702165}}].

\bibitem{Sen:1996vd}
A.~Sen, {\it {F-theory and Orientifolds}},  {\em Nucl. Phys.} {\bf B475} (1996)
  562--578, [\href{http://arxiv.org/abs/hep-th/9605150}{{\tt hep-th/9605150}}].

\bibitem{Sen:1997kw}
A.~Sen, {\it {F-theory and the Gimon-Polchinski orientifold}},  {\em Nucl.
  Phys.} {\bf B498} (1997) 135--155,
  [\href{http://arxiv.org/abs/hep-th/9702061}{{\tt hep-th/9702061}}].

\bibitem{Sethi:1996es}
S.~Sethi, C.~Vafa, and E.~Witten, {\it {Constraints on low-dimensional string
  compactifications}},  {\em Nucl. Phys.} {\bf B480} (1996) 213--224,
  [\href{http://arxiv.org/abs/hep-th/9606122}{{\tt hep-th/9606122}}].

\bibitem{Witten:1996bn}
E.~Witten, {\it {Non-Perturbative Superpotentials In String Theory}},  {\em
  Nucl. Phys.} {\bf B474} (1996) 343--360,
  [\href{http://arxiv.org/abs/hep-th/9604030}{{\tt hep-th/9604030}}].

\bibitem{Denef:2008wq}
F.~Denef, {\it {Les Houches Lectures on Constructing String Vacua}},
  \href{http://arxiv.org/abs/0803.1194}{{\tt arXiv:0803.1194}}.

\bibitem{Denef:2004dm}
F.~Denef, M.~R. Douglas, and B.~Florea, {\it {Building a better racetrack}},
  {\em JHEP} {\bf 06} (2004) 034,
  [\href{http://arxiv.org/abs/hep-th/0404257}{{\tt hep-th/0404257}}].

\bibitem{Giryavets:2003vd}
A.~Giryavets, S.~Kachru, P.~K. Tripathy, and S.~P. Trivedi, {\it {Flux
  compactifications on Calabi-Yau threefolds}},  {\em JHEP} {\bf 04} (2004)
  003, [\href{http://arxiv.org/abs/hep-th/0312104}{{\tt hep-th/0312104}}].

\bibitem{Klemm:1996ts}
A.~Klemm, B.~Lian, S.~S. Roan, and S.-T. Yau, {\it {Calabi-Yau fourfolds for M-
  and F-theory compactifications}},  {\em Nucl. Phys.} {\bf B518} (1998)
  515--574, [\href{http://arxiv.org/abs/hep-th/9701023}{{\tt hep-th/9701023}}].

\bibitem{Denef:2005mm}
F.~Denef, M.~R. Douglas, B.~Florea, A.~Grassi, and S.~Kachru, {\it {Fixing all
  moduli in a simple F-theory compactification}},  {\em Adv. Theor. Math.
  Phys.} {\bf 9} (2005) 861--929,
  [\href{http://arxiv.org/abs/hep-th/0503124}{{\tt hep-th/0503124}}].

\bibitem{Dasgupta:1997cd}
K.~Dasgupta, D.~P. Jatkar, and S.~Mukhi, {\it {Gravitational couplings and Z(2)
  orientifolds}},  {\em Nucl. Phys.} {\bf B523} (1998) 465--484,
  [\href{http://arxiv.org/abs/hep-th/9707224}{{\tt hep-th/9707224}}].

\bibitem{Collinucci:2008pf}
A.~Collinucci, F.~Denef, and M.~Esole, {\it {D-brane Deconstructions in IIB
  Orientifolds}},  \href{http://arxiv.org/abs/0805.1573}{{\tt
  arXiv:0805.1573}}.

\bibitem{Sen:1998sm}
A.~Sen, {\it {Tachyon condensation on the brane antibrane system}},  {\em JHEP}
  {\bf 08} (1998) 012, [\href{http://arxiv.org/abs/hep-th/9805170}{{\tt
  hep-th/9805170}}].

\bibitem{Uranga:2000xp}
A.~M. Uranga, {\it {D-brane probes, RR tadpole cancellation and K-theory
  charge}},  {\em Nucl. Phys.} {\bf B598} (2001) 225--246,
  [\href{http://arxiv.org/abs/hep-th/0011048}{{\tt hep-th/0011048}}].

\bibitem{Minasian:1997mm}
R.~Minasian and G.~W. Moore, {\it {K-theory and Ramond-Ramond charge}},  {\em
  JHEP} {\bf 11} (1997) 002, [\href{http://arxiv.org/abs/hep-th/9710230}{{\tt
  hep-th/9710230}}].

\bibitem{Scrucca:1999uz}
C.~A. Scrucca and M.~Serone, {\it {Anomalies and inflow on D-branes and
  O-planes}},  {\em Nucl. Phys.} {\bf B556} (1999) 197--221,
  [\href{http://arxiv.org/abs/hep-th/9903145}{{\tt hep-th/9903145}}].

\bibitem{Stefanski:1998yx}
J.~Stefanski, Bogdan, {\it {Gravitational couplings of D-branes and O-planes}},
   {\em Nucl. Phys.} {\bf B548} (1999) 275--290,
  [\href{http://arxiv.org/abs/hep-th/9812088}{{\tt hep-th/9812088}}].

\bibitem{Morales:1998ux}
J.~F. Morales, C.~A. Scrucca, and M.~Serone, {\it {Anomalous couplings for
  D-branes and O-planes}},  {\em Nucl. Phys.} {\bf B552} (1999) 291--315,
  [\href{http://arxiv.org/abs/hep-th/9812071}{{\tt hep-th/9812071}}].

\bibitem{Anderson:2008uw}
L.~B. Anderson, Y.-H. He, and A.~Lukas, {\it {Monad Bundles in Heterotic String
  Compactifications}},  {\em JHEP} {\bf 07} (2008) 104,
  [\href{http://arxiv.org/abs/0805.2875}{{\tt arXiv:0805.2875}}].

\end{thebibliography}\endgroup

\end{document}